\documentclass{article}
\usepackage{amsfonts}

\usepackage{graphicx}
\usepackage{amsmath}

\newenvironment{proof}[1][Proof]{\textbf{#1.} }{\ \rule{0.5em}{0.5em}}
\input{tcilatex}

\begin{document}

\title{THE STOCHASTIC SECTOR OF INTERACTING-FREE QUANTUM FIELD THEORY}
\author{John Gough \\
Department of Computing \& Mathematics\\
Nottingham-Trent University, Burton Street,\\
Nottingham NG1\ 4BU, United Kingdom.\\
john.gough@ntu.ac.uk}
\date{}
\maketitle

\begin{abstract}
The Quantum Stochastic Limit of a quantum mechanical particle coupled to a
quantum field without the neglect of the response details of the interaction
(i.e. not making the dipole approximation) is made following the treatment
of Accardi and Lu [6] and the corresponding Quantum Stochastic Structure is
derived. The stochastic sector for the noise is constructed and is shown to
be of a qualitatively new type. We also include a physical discussion on the
limit noise which obeys Interacting-Free statistics and include a new
shorter proof of the noise convergence and also a new construction of
Interacting-Free Fock Space.
\end{abstract}

\section{Introduction.}

The theory of stochastic processes has many deep connections with quantum
field theory. The path integral approach of Feynman [1], in particular,
reveals close analogies between quantum field theory in real time and
Brownian motion. An important line of research which has deepened this
connection in recent years is that of quantum stochastic approximations:
here one considers a test system $S$ (quantum mechanical) coupled to an
infinite reservoir $R$ (a Bosonic quantum field), the Hamiltonian for the
combined system and reservoir takes the form $H=H_{S}+H_{R}+\lambda H_{I}$
where only the interaction $H_{I}$ couples S to R. A Gaussian state (e.g.
vacuum or thermal) is prescribed for the reservoir and one makes a
separation of time scales (van Hove limit): time $t$ being rescaled as $%
t/\lambda ^{2}$ followed by the limit $\lambda \rightarrow 0$. In an
approach pioneered by Accardi, Frigerio and Lu [2], one constructs suitable
collective reservoir fields in which to examine the limiting behaviour of
observables and these collective fields have the property of themselves
converging to basic quantum stochastic processes (typically quantum Brownian
motion). This fact was exploited by Accardi, Lu and Volovich [3] to
establish a (quantum) stochastic sector in quantum field theory.

The original scope of [2] was very limited due to the fact that almost all
the standard simplifying assumptions (vacuum state, rotating wave
approximation, dipole approximation, etc.) were made in order to make an
already complicated problem accessible. However, since then, these
assumptions have been removed with relative ease [4]. The connection between
the quantum stochastic limit theory (when applied to an atomic system of
bound states: i.e. when $H_S$ has discrete spectrum) and the standard
application of the \textit{Golden Rule} to the same problem has been
explained in Accardi, Gough and Lu [5].

Recently the problem of considering a system with continuous spectrum
without recourse to the dipole approximation has been tackled [6]. The
surprising feature which emerges is that, by now including all the details
of the interaction between $S$ and $R$, the limit quantum noise has a
qualitatively new character. Instead of inheriting the Bose statistics of $R$%
, the noise in fact obeys a non-linear modification of the Free statistics.
The originally notion of Free-ness is due to Voiculescu [7] and in the
context of quantum stochastic theory was first studied by K\"{u}mmerer and
Speicher [8]. We shall use the term \emph{Interacting-Free} to describe the
noise studied here: the notion of Interacting Fock space over a Fock Module
necessary to describe the limit noise was introduced however by Lu [9].

The goal of this paper is to extend the notion of (quantum) stochastic
sector so that the interacting-free field limit can be included.

\subsection{The Physical Model}

As system we consider a quantum mechanical particle with spin zero and
unperturbed Hamiltonian $H_{S}$: 
\begin{equation}
H_{S}=\frac{p^{2}}{2m}.  \tag{$1.1$}
\end{equation}
Here $p$ is canonical momentum with canonical position denoted by $q$: $%
[q_{j},p_{l}]=i\hbar \delta _{j,l}$.

The reservoir is taken, for transparency, to have spinless Bosonic quanta.
We denote by $a^{\dagger }(k)$ the creation operator for a reservoir quantum
of momentum $k$. Along with its adjoint $a(k)$ we have the canonical
commutation relations 
\begin{equation}
\lbrack a(k),a^{\dagger }(k^{\prime })]=\delta (k-k^{\prime }).  \tag{$1.2$}
\end{equation}
The unperturbed Hamiltonian for the reservoir is taken to be 
\begin{equation}
H_{R}=\int dk\,\hbar \omega (k)a^{\dagger }(k)a(k),  \tag{$1.3$}
\end{equation}
where $\omega (k)\geq 0$ gives the dispersion relation for $R$.

The unperturbed evolution operator for $S+R$ is then 
\begin{equation}
V_{t}^{0}=\exp \{{\frac{t}{i\hbar }}(H_{S}\otimes 1_{R}+1_{S}\otimes
H_{R})\}.  \tag{$1.4$}
\end{equation}
The interaction between the particle and field takes the form 
\begin{equation}
H_{I}=D(p)\mathcal{A}(q)  \tag{$1.5$}
\end{equation}
where $D(p)$ is an observable of the system and 
\begin{equation}
\mathcal{A}(q)=\int dk\,\{g(k)e^{-ik.q}\otimes a^{\dagger }(k)+\overline{g}%
(k)e^{ik.q}\otimes a(k)\}.  \tag{$1.6$}
\end{equation}
$\mathcal{A}(q)$ is the potential of the field and naturally depends on the
particle's position $q$. The form factor $g$ is taken to be a Schwartz
function on $\mathbb{R}$. We shall assume that $[D(p),\mathcal{A}(q)]=0$ so
that $H_{I}$ is self adjoint. For the situation of an electron coupled to
the QED field, the reservoir quanta (photons) have polarization and we can
choose the Coulomb gauge so that $D(p)\equiv -{\frac{e}{m}}p$ commutes with $%
\mathcal{A}(q)$: in this case we would of course have a vector product. In
our case, in order to study the field in detail with the only simplifying
assumption that the quanta be spinless, we make the assumption that $D$ is
proportional to $1_{S}$ and drop it entirely. This in fact changes very
little in the qualitative description of the limiting noise.

We remark that the reverse situation is considered in most other treatments:
that is, one assumes that $D$ is $p$-dependent while $\mathcal{A}$ is $q$%
-independent. In such cases, we say that the field is \emph{responseless}:
then there is the replacement 
\begin{equation}
\mathcal{A}\mapsto \mathcal{A}^{\prime }=\int dk\,\{g(k)a^{\dagger }(k)+%
\overline{g}(k)a(k)\}.  \tag{$1.7$}
\end{equation}
In the QED case this is the dipole approximation. Now $\mathcal{A}^{\prime }$
is trivial and a test system S, under this replacement, cannot obtain any
measurable information about the individual modes of the field (because this
is precisely the detail which is elided in $\mathcal{A}^{\prime }$). The
situation of a responseless field has already been studied in the quantum
stochastic limit and it is known that a quantum Brownian motion emerges.

Our objective is then to make a study of the \emph{responsive} interaction 
\begin{equation}
H_{I}=c^{\dagger }(g)+c(g),  \tag{$1.8$}
\end{equation}
where we introduce the combined (interacting) fields 
\begin{equation}
c^{\dagger }(g)=\int dk\,g(k)e^{-ik.q}\otimes a^{\dagger }(k),\,c(g)=\int
dk\,\overline{g}(k)e^{ik.q}\otimes a(k).  \tag{$1.9$}
\end{equation}
The total Hamiltonian is taken to be 
\begin{equation}
H_{\lambda }=\{H_{S}\otimes 1_{R}+1_{S}\otimes H_{R}\}+\lambda H_{I} 
\tag{$1.10$}
\end{equation}
where $\lambda $ is a non-zero coupling constant.

The van Hove scaling limit has, in previous applications to quantum
stochastic limits, suggested the use of collective operator fields of the
following type 
\begin{equation}
c_{t,\lambda }^{\sharp }(g):=\lambda \int_{0}^{t/\lambda ^{2}}d\tau
\,V_{\tau }^{0\dagger }c^{\sharp }(g)V_{\tau }^{0}.  \tag{$1.11$}
\end{equation}
The limit of such collective operators in the vacuum field $\Psi _{R}$ of
the reservoir was obtained in [6]. The limiting fields, denoted by $%
C^{\sharp }(g,t)$, do not satisfy Bose commutation relations but, on account
of the response factor $\exp \{\mp ik.q\}$ which couples the system to all
modes of the field, satisfy a modified version of the free relations.

The Weyl operators offer a straightforward means to study the unperturbed
evolution of the response factors so we review them now. For $a,b\in \mathbb{%
R}^{3}$, we define the Weyl operator $W(a,b)$ to be the unitary operator 
\begin{equation}
W(a,b)=e^{i(a.p+b.q)}.  \tag{$1.12$}
\end{equation}
They satisfy

\begin{itemize}
\item[1)]  $W(a,b)=e^{ia.p}e^{ib.q}e^{-i\hbar
a.b/2}=e^{ib.q}e^{ia.p}e^{i\hbar a.b/2};$

\item[2)]  $W(a_{1},b_{1})W(a_{2},b_{2})=W(a_{1}+a_{2},b_{1}+b_{2})\exp \{{%
\frac{i\hbar }{2}}(a_{1}.b_{2}-a_{2}.b_{1})\}$ or, more generally, 
\begin{equation*}
W(a_{1},b_{1})...W(a_{n},b_{n})=W(\sum_{j}a_{j},\sum_{j}b_{j})\exp \{{\frac{%
i\hbar }{2}}\sum_{j<l}(a_{j}.b_{l}-a_{l}.b_{j})\};
\end{equation*}

\item[3)]  $W(a,b)^{\dagger }=W(-a,-b);$

\item[4)]  Under the unperturbed evolution we have $p_{t}=p,q_{t}=q+{\frac{t%
}{m}}p$ and so the Weyl operators evolve as shown below 
\begin{equation*}
W(a,b)_{t}=e^{i(a.p_{t}+b.q_{t})}=e^{i((a+{\frac{t}{m}})p+b.q)}=W(a+{\frac{t%
}{m}}b,b).
\end{equation*}
\end{itemize}

\bigskip

Therefore we have 
\begin{equation}
V_{\tau }^{o\dagger }c^{\dagger }(g)V_{\tau }^{0}=\int dk\,g(k)e^{i\omega
(k)\tau }W(-{\frac{\tau }{m}}k,-k)\otimes a^{\dagger }(k).  \tag{$1.13$}
\end{equation}

\section{The Quantum Stochastic Sector}

The results of Accardi, Lu and Volovich [3] can be summarized as follows:-
for $\omega >0$ let 
\begin{equation}
B_{t,\lambda }^{\dagger }(g):=\lambda \int_{0}^{t/\lambda ^{2}}d\tau \int
dk\,g(k)e^{i[\omega (k)-\omega ]\tau }a^{\dagger }(k).  \tag{$2.1$}
\end{equation}
which is the collective operator describing a responseless field. As $%
\lambda \rightarrow 0$ one shows that $B_{t,\lambda }^{\sharp }(g)$
converges to Bosonic quantum Brownian motion $B^{\sharp }(g,t)$ satisfying 
\begin{equation}
\lbrack B(g,t),B^{\dagger }(f,s)]=(t\wedge s)\,(g|f)  \tag{$2.2$}
\end{equation}
where 
\begin{equation}
(g|f):=\int_{-\infty }^{+\infty }d\tau \int dk\,e^{i[\omega (k)-\omega ]\tau
}\overline{g}(k)f(k)\equiv \int dk\,\delta (\omega (k)-\omega )\overline{g}%
(k)f(k).  \tag{$2.3$}
\end{equation}
Now $\omega $ can be interpreted as a probing frequency: that is $\omega $
is a frequency associated to $D$ under rotating wave approximation and in
principle different test systems (having different resonant $\omega $)
reveal further information about the reservoir. However in this case the
detailed information is restricted by the responseless assumption.

The noise fields $b_{\lambda }^{\sharp }(u,k)$ defined by 
\begin{equation}
b_{\lambda }^{\dagger }(u,k):={\frac{1}{\lambda }}e^{i\omega u/\lambda
^{2}}a^{\dagger }(k)  \tag{$2.4$}
\end{equation}
so that 
\begin{equation}
B_{t,\lambda }^{\sharp }(g)\equiv \int_{0}^{t}du\int dk\,g(k)b^{\sharp
}(u,k),  \tag{$2.5$}
\end{equation}
then converge in the limit $\lambda \rightarrow 0$ in the vacuum state to
the quantum white noise $b^{\sharp }(u,k)$ satisfying 
\begin{equation}
\lbrack b(u,k),b^{\dagger }(u^{\prime },k^{\prime })]=2\pi \delta
(u-u^{\prime })\delta (\omega (k)-\omega )\delta (k-k^{\prime }). 
\tag{$2.6$}
\end{equation}
The results of this paper can then be summarized as follows. Introducing the
density operators 
\begin{equation}
a_{\lambda }^{\dagger }(u,k):={\frac{1}{\lambda }}V_{u/\lambda
^{2}}^{0\dagger }[e^{-ik.q}\otimes a^{\dagger }(k)]V_{u/\lambda ^{2}}^{0} 
\tag{$2.7$}
\end{equation}
we have the limit (in law) $a_{\lambda }^{\dagger }(u,k)\rightarrow
a^{\dagger }(u,k)$ where 
\begin{equation}
a^{\dagger }(u,k)=\int_{-\infty }^{+\infty }d\tau \,e^{i\omega (k)\tau }W(-{%
\frac{\tau }{m}}k,-k)\otimes a^{\dagger }(u,\tau ,k)  \tag{$2.8$}
\end{equation}
and $a^{\sharp }(u,\tau ,k)$ satisfy the \emph{modified Free relations} 
\begin{equation}
a(u,\tau ,k)a^{\dagger }(u^{\prime },\tau ^{\prime },k^{\prime })=\delta
(u-u^{\prime })\delta (2\tau -\tau ^{\prime })\delta (k-k^{\prime }). 
\tag{$2.9$}
\end{equation}
The limiting collective operator $C^{\dagger }(g,t)$ may then be expressed
as 
\begin{equation}
C^{\dagger }(g,t)=\int_{0}^{t}du\int_{-\infty }^{+\infty }d\tau \int
dk\,g(k)e^{i\omega (k)\tau }W(-{\frac{\tau }{m}}k,-k)\otimes a^{\dagger
}(u,\tau ,k).  \tag{$2.10$}
\end{equation}
Note: we can introduce fields $\alpha ^{\sharp }(\tau ,k)$ satisfying the
relations 
\begin{equation}
\alpha (\tau ,k)\alpha ^{\dagger }(\tau ^{\prime },k^{\prime })=\delta
(2\tau -\tau ^{\prime })\delta (k-k^{\prime }).  \tag{$2.11$}
\end{equation}
and set 
\begin{eqnarray}
C^{\dagger }(g,t) &=&|\chi _{\lbrack 0,t]}>\underline{\otimes }\int_{-\infty
}^{+\infty }\int dk\,g(k)e^{i\omega (k)\tau }W(-{\frac{\tau }{m}}%
k,-k)\otimes \alpha ^{\dagger }(\tau ,k);  \notag \\
C(g,t) &=&<\chi _{\lbrack 0,t]}|\underline{\otimes }\int_{-\infty }^{+\infty
}\int dk\,\overline{g}(k)e^{-i\omega (k)\tau }W({\frac{\tau }{m}}k,k)\otimes
\alpha (\tau ,k)  \TCItag{$2.12$}
\end{eqnarray}
where $|\alpha >$ is -ket and $<\alpha |$ is bra- for $\alpha \in \mathcal{L}%
^{2}(\mathbb{R})$. If we have an ordered product of $C^{\sharp
}(g_{j},t_{j}) $ then we will have an associated ordered product of bras and
kets: the simple algebraic rule applied here is that whenever a bra is
immediately to the left of a ket they form a scalar product and can be taken
to one side. Thus, for instance, 
\begin{equation*}
<\alpha |\,.\,|\beta >\,:=<\alpha ,\beta >\equiv \int_{-\infty }^{+\infty
}dt\,\overline{\alpha }(t)\beta (t),
\end{equation*}
\begin{equation*}
<\alpha _{1}|\,.\,|\beta _{1}>\,.\,<\alpha _{2}|\,.\,|\beta _{2}>=<\alpha
_{1},\beta _{1}><\alpha _{2},\beta _{2}>
\end{equation*}
while 
\begin{equation*}
<\alpha _{1}|\,.\,<\alpha _{2}|\,.\,|\beta _{2}>\,.\,|\beta _{1}>=<\alpha
_{1},\beta _{1}><\alpha _{2},\beta _{2}>
\end{equation*}
It is the compatibility of the bra-ket formalism with the free statistics
that allows the description of $C^{\dagger }(g,t)$ as algebraic tensor
product of $\mathcal{L}^{2}(\mathbb{R})$ with the $W\otimes \alpha ^{\sharp
} $-operators (c.f. remark b in section 3).

\section{The Limit Processes, \textbf{$C^{\sharp }(g,t)$}}

In the following we shall adopt the convention that $%
\prod_{j=1}^{n}X_{j}=X_{n}...X_{1}$ and that, for any operator $X$, $%
X^{0}:=X $ while $X^{1}:=X^{\dagger }$. A sequence $\varepsilon =\{\epsilon
_{2n},...,\epsilon _{1}\}\in \{0,1\}^{2n}$ will be referred to as \emph{%
non-trivial }if $<\Psi _{R},\prod_{j=1}^{2n}a^{\epsilon _{j}}(k_{j})\Psi
_{R}>$ is not identically zero. Clearly, there must be an equal number of
creators and annihilators if the expectation above is to be non-zero.
Suppose $\varepsilon $ is non-trivial and let $M=(m_{n},...,m_{1})$ denote
the set of creator indices (i.e. $\epsilon _{m_{j}}=1)$ ordered so that $%
m_{h}<m_{h+1}$. Let $M^{c}$ then denote the (unordered) set of annihilator
index positions. To guarantee non-triviality we also require the condition
that for all $r=1,...,2n$ 
\begin{equation}
\sharp \{m^{\prime }\in M^{c}:m^{\prime }\leq r\}\leq \max \{h:m_{h}\leq r\}.
\tag{$3.1$}
\end{equation}
The odd correlation functions clearly vanish in the reservoir vacuum. The
even correlators are given below. They where first computed by Accardi and
Lu [6], however we include here a shorter proof (Theorem 1).

Before this we give a brief account of Free statistics. Let $T$ be a space
of test functions with inner product $\langle .,.\rangle $. Let $b(g)$, $%
b^{\dagger }(g)$ be operators (for each $g\in T$) and $\Psi $ a vector such
that 
\begin{equation}
b(g)^{\dagger }=b^{\dagger }(g)\ ,\,b(g)\Psi =0  \tag{$3.2$}
\end{equation}
and 
\begin{equation}
b(g)b^{\dagger }(f)=\langle g,f\rangle  \tag{$3.3$}
\end{equation}
for all $g$, $f\in T$. The operators $b^{\sharp }(g)$ are said to satisfy 
\emph{free statistics} or \emph{free relations}. $\Psi $ is referred to as
the \emph{Fock vacuum vector}. An explicit construction can be given on Fock
space $\Gamma (T)=\oplus _{n=1}^{\infty }(\otimes ^{n}T)$ over $T$ by taking 
$b^{\dagger }(g)$ to be the mapping $:\phi _{1}\otimes \dots \otimes \phi
_{n}\mapsto g\otimes \phi _{1}\otimes \dots \otimes \phi _{n}$. In which
case, its adjoint is $b(g):\phi _{1}\otimes \dots \otimes \phi _{n}\mapsto
\langle g,\phi _{1}\rangle \phi _{2}\otimes \dots \otimes \phi _{n}$. The
vacuum vector is then the Fock vacuum $\Psi :=1\oplus 0\oplus 0\dots $.

Now it is easily seen that $\langle \Psi ,\prod_{j}^{2n}b^{\epsilon
_{j}}(g_{j})\Psi \rangle $ is not identically zero provided $\varepsilon $
is again non-trivial. However the relations (3.3) give that 
\begin{equation}
\langle \Psi ,\prod_{j}^{2n}b^{\epsilon _{j}}(g_{j})\Psi \rangle
=\prod_{j}^{n}\langle g_{\overline{m}_{j}},g_{m_{j}}\rangle ,  \tag{$3.4$}
\end{equation}
where $(\overline{m}_{n},...,\overline{m}_{1})$ is the unique ordered
sequence which agrees with $M^{c}$ as a set and satisfies

\begin{itemize}
\item[i)]  $\overline{m}_{j}>m_{j}$

\item[ii)]  $\forall \,h=1,\dots ,n$ 
\begin{equation}
\overline{m}_{h}>m_{j}>m_{h}\Leftrightarrow \overline{m}_{h}>\overline{m}%
_{j}>m_{h}  \tag{$3.5$}
\end{equation}
\end{itemize}

\textbf{Remark:}\textit{\ }Condition (i) comes from having to arrange $%
\prod_{j=1}^{2n}b^{j}(g_{j})$ in normal order. Note that the logical
negation of (ii) also holds, that is if $m_{j}$ lies outside of $\{\overline{%
m}_{n},\dots ,m_{n}\}$ then so too does $\overline{m}_{j}$, and \textit{vice
versa}. The set of $n$ pairs $\{(\overline{m}_{n},m_{n}):h=1,\dots ,n\}$ as
above is called the \emph{Wigner} or \emph{non-crossing} pair partition of $%
\varepsilon $. \bigskip

\textbf{Remark:} We have already met an example of freeness in our algebraic
rule for bras and kets in the last section: the identification 
\begin{equation}
|\alpha >\equiv b^{\dagger }(\alpha ),\,<\beta |\equiv b(\beta ), 
\tag{$3.6$}
\end{equation}
for $T\equiv L^{2}(\mathbb{R})$, now makes this rule definite.

\bigskip

\noindent \textbf{Theorem 1}. \emph{Let }$\varepsilon \in \{0,1\}^{2n}$\emph{%
\ be non-trivial then } 
\begin{eqnarray}
\langle \prod_{j=1}^{2n}C^{\epsilon _{j}}(g_{j},T_{j})\rangle
&:&=\lim_{\lambda \rightarrow 0}<\Psi _{R},\prod_{j=1}^{2n}c_{T_{j},\lambda
}^{\epsilon _{j}}(g_{j})\Psi _{R}>  \TCItag{$3.7$} \\
&=&\prod_{j=1}^{n}(T_{\overline{m}_{j}}\wedge T_{m_{j}})\int_{-\infty
}^{\infty }d\tau _{m_{1}}\dots \int_{-\infty }^{\infty }d\tau _{m_{n}} 
\notag
\end{eqnarray}
\begin{equation*}
\times \int d^{3}k_{1}\dots \int d^{3}k_{n}\,\prod_{h=1}^{n}\{\overline{g}_{%
\overline{m}_{h}}(k_{m_{h}})s_{m_{h}}\exp \{i[\omega (k_{m_{h}}+{\frac{\hbar 
}{2m}}|k_{m_{h}}|^{2}]\tau _{m_{h}}\}
\end{equation*}
\begin{equation}
\times \exp \{-{\frac{i}{m}}p.\sum_{j=1}^{n}k_{m_{j}}\tau _{m_{j}}\}\,\exp \{%
{\frac{i\hbar }{m}}\sum_{h,r=1}^{n}\tau _{m_{h}}k_{m_{h}}.k_{m_{r}}\chi _{(%
\overline{m}_{r},m_{r})}(m_{h})\}  \tag{$3.8$}
\end{equation}
\emph{where }$\{(\overline{m}_{j},m_{j}):j=1,\dots ,n\}$\emph{\ is the
Wigner (non-crossing) partition of }$\{1,\dots ,2n\}$\emph{\ associated with 
}$\varepsilon $\emph{. If }$\varepsilon $\emph{\ is trivial then (3.7)
vanishes. }

The origin of this limit can be explained as follows. In principle the $2n$%
-point correlations before the limit can be expressed (due to the Bosonic
nature of the reservoir and our choice of a Gaussian state) in terms of all
possible pair partitions. Now retaining the response term means that for
each emission and absorption of a reservoir quantum we keep the details of
the momentum recoil of the system particle, and so enforcing strict momentum
conservation. A contracted creation and annihilation pair survives the
stochastic limit only if it is energetically balanced: this amounts to the
Golden rule. However the only complete set of pair partitions which has all
contracted pairs energetically balanced (and here we must have momentum
conservation) is the Wigner pair partition, if one exists.

\bigskip

\begin{proof}
For $\epsilon \in \{1,0\}$, we have 
\begin{equation}
c_{T,\lambda }^{\epsilon }(g)\equiv \lambda \int_{0}^{T/\lambda ^{2}}d\tau
\int dk\,g^{\epsilon }(k)\exp \{i(-1)^{\epsilon }\omega (k)\tau \}W({\frac{%
(-1)^{\epsilon }\tau }{m}}k,(-1)^{\epsilon }k)\otimes a^{\epsilon }(k). 
\tag{$3.9$}
\end{equation}
Here we set 
\begin{equation}
g^{0}(k)=\overline{g}(k),\,g^{1}(k)=g(k).  \tag{$3.10$}
\end{equation}
For $\varepsilon =\{\epsilon _{2n},...,\epsilon _{1}\}\in \{1,0\}^{2n}$
non-trivial, we have 
\begin{equation*}
<\Psi _{R},\prod_{j=1}^{2n}c_{T_{j},\lambda }^{\epsilon _{j}}(g_{j})\Psi
_{R}>=\lambda ^{2n}\prod_{j=1}^{2n}\left\{ \int_{0}^{T_{j}/\lambda
^{2}}d\tau _{j}\int d^{3}k_{j}\,g_{j}^{\epsilon _{j}}(k_{j})\exp
\{i(-1)^{\epsilon _{j}}\omega (k_{j})\tau _{j}\}\right\}
\end{equation*}
\begin{equation}
\times \prod_{l=1}^{2n}W({\frac{(-1)^{\epsilon _{j}}\tau _{j}}{m}}%
k_{j},(-1)^{\epsilon _{j}}k_{j})<\Psi _{R},\prod_{h=1}^{2n}a^{\epsilon
_{h}}(k_{h})\Psi _{R}>.  \tag{3.11}
\end{equation}
but 
\begin{equation}
<\Psi _{R},\prod_{h=1}^{2n}a^{\epsilon _{n}}(k_{h})\Psi
_{R}>=\sum_{\{M^{\prime }\equiv M^{c}:m_{n}^{\prime }<m_{h}\forall
\,h\}}\prod_{h=1}^{n}\delta (k_{m_{n}^{\prime }}-k_{m_{n}})  \tag{$3.12$}
\end{equation}
that is, we sum over all possible pair contractions of creator--annihilator
indices $\{(m_{h}^{\prime },m_{h}):\,\,\,h=1,\dots ,n\}$ where $M^{\prime
}=(m_{n}^{\prime },...,m_{1}^{\prime })$ is equivalent to $M^{c}$ as a set.
As we produce contractions by moving terms to normal order, we clearly need
only consider $m_{h}^{\prime }>m_{h}$ however: that is to say the pair
contraction $(m_{h}^{\prime },m_{h})$ only comes about from having to move
the annihilator $a(k_{m_{h}^{\prime }})$ from the left to the right of $%
a^{\dagger }(k_{m_{h}})$.

Therefore we may write 
\begin{equation*}
<\Psi _{R},\prod_{j=1}^{2n}c_{T_{j},\lambda }^{\epsilon _{j}}(g_{j})\Psi
_{R}>=
\end{equation*}
\begin{equation*}
\sum_{\{M^{\prime }\equiv M^{c}:m_{n}^{\prime }<m_{h}\forall
\,h\}}\prod_{h=1}^{n}\biggl\{\lambda ^{2}\int_{0}^{T_{{m^{\prime }}%
_{h}/\lambda ^{2}}}d\tau _{{m^{\prime }}_{h}}\int_{0}^{T_{m_{h}}/\lambda
^{2}}d\tau _{m_{h}}\int dk_{m_{h}}
\end{equation*}
\begin{equation}
\overline{g}_{{m^{\prime }}_{h}}(k_{m_{h}})g_{m_{h}}(k_{m_{h}})\exp
\{i\omega (k_{m_{h}})\tau _{m_{h}}\}\biggr\}\prod_{l=1}^{2n}W({\frac{%
(-1)^{\epsilon _{l}}\tau _{l}}{m}}k_{l},(-1)^{\epsilon _{l}}k_{l}), 
\tag{$3.13$}
\end{equation}
where the product of Weyl operators must be accompanied by the relevant
assignment $k_{m_{j}}=k_{{m^{\prime }}_{j}}$ for each ${M^{\prime }}$
considered in the sum.

Now, using the rule for multiplying Weyl operators and mindful of our
product convention, we have that 
\begin{equation*}
\prod_{l=1}^{2n}W\left( (-1)^{\epsilon _{l}}{\frac{\tau _{l}}{m}}%
k_{l},\,(-1)^{\epsilon _{l}}k_{l}\right) =\exp \{-{\frac{i\hbar }{2m}}%
\sum_{1\leq j<l\leq 2n}(-1)^{\epsilon _{j}+\epsilon _{l}}k_{j}.k_{l}(\tau
_{j}-\tau _{l})\}
\end{equation*}
\begin{equation}
\times W\left( \sum_{1\leq l\leq 2n}(-1)^{\epsilon _{l}}{\frac{\tau _{l}}{m}}%
\,k_{l}\ ,\,\sum_{1\leq l\leq 2n}(-1)^{\epsilon _{l}}k_{l}\right) 
\tag{$3.14$}
\end{equation}
Momentum balance requires that 
\begin{equation}
\sum_{1\leq l\leq 2n}(-1)^{\epsilon _{l}}k_{l}=0,  \tag{$3.15$}
\end{equation}
so the correlation function is independent of $q$ and so diagonal in $p$. 
\begin{equation}
\sum_{1\leq l\leq 2n}(-1)^{\epsilon _{l}}{\frac{\tau _{l}}{m}}k_{l}=-{\frac{1%
}{m}}\sum_{1\leq h\leq n}(\tau _{m_{h}}-\tau _{{m^{\prime }}_{h}})k_{m_{h}}.
\tag{$3.16$}
\end{equation}
The phase associated with ${M^{\prime }}$ is then 
\begin{equation*}
{\frac{-i\hbar }{2m}}\sum_{l=1}^{2n}\sum_{j<l}(-1)^{\epsilon _{j}+\epsilon
_{l}}k_{j}.k_{l}(\tau _{j}-\tau _{l})
\end{equation*}
\begin{equation*}
={\frac{-i\hbar }{2m}}\sum_{h=1}^{n}\left\{ \sum_{1\leq j<{m^{\prime }}%
_{h}}(-1)^{\epsilon _{j}}k_{j}.k_{{m^{\prime }}_{h}}(\tau _{j}-\tau _{{%
m^{\prime }}_{h}})-\sum_{1\leq j<m_{h}}(-1)^{\epsilon
_{j}}k_{j}.k_{m_{h}}(\tau _{j}-\tau _{m_{h}})\right\}
\end{equation*}
\begin{equation*}
={\frac{-i\hbar }{2m}}\sum_{h=1}^{n}\biggl\{\sum_{\alpha }^{{m^{\prime }}%
_{\alpha }<{m^{\prime }}_{h}}k_{{m^{\prime }}_{\alpha }}.k_{{m^{\prime }}%
_{h}}(\tau _{{m^{\prime }}_{\alpha }}-\tau _{{m^{\prime }}_{h}})-\sum_{\beta
}^{m_{\beta }<{m^{\prime }}_{h}}k_{m_{\beta }}.k_{{m^{\prime }}_{h}}(\tau
_{m_{\beta }}-\tau _{{m^{\prime }}_{h}})
\end{equation*}
\begin{equation}
-\sum_{\gamma }^{{m^{\prime }}_{\gamma }<m_{h}}k_{{m^{\prime }}_{\gamma
}}.k_{m_{h}}(\tau _{{m^{\prime }}_{\gamma }}-\tau _{m_{h}})+\sum_{\delta
}^{m_{\delta }<m_{h}}k_{m_{\delta }}.k_{m_{h}}(\tau _{m_{\delta }}-\tau
_{m_{h}})\biggr\}  \tag{$3.17a$}
\end{equation}
and putting together the first term with the third and second with fourth 
\begin{equation*}
={\frac{-i\hbar }{2m}}\sum_{h=1}^{n}\biggl\{\sum_{\alpha }^{{m^{\prime }}%
_{\alpha }<{m}_{h}}k_{m_{\alpha }}.k_{m_{h}}(\tau _{m_{h}}-\tau _{{m^{\prime
}}_{h}})-\sum_{\beta }^{m_{\beta }<{m}_{h}}k_{m_{\beta }}.k_{m_{h}}(\tau
_{m_{h}}-\tau _{{m^{\prime }}_{h}})
\end{equation*}
\begin{equation*}
-\sum_{\alpha }^{{m}_{h}<{m^{\prime }}_{\alpha }<{m^{\prime }}%
_{h}}k_{m_{\alpha }}.k_{m_{h}}(\tau _{{m^{\prime }}_{\alpha }}-\tau
_{m_{h}})-\sum_{\beta }^{{m}_{h}<m_{\beta }<{m^{\prime }}_{h}}k_{m_{\beta
}}.k_{m_{h}}(\tau _{m_{\beta }}-\tau _{{m^{\prime }}_{h}})
\end{equation*}
\begin{equation}
-k_{m_{h}}.k_{m_{h}}(\tau _{m_{h}}-\tau _{{m^{\prime }}_{h}})\biggr\}. 
\tag{$3.17b$}
\end{equation}
We now undergo a chance of variables 
\begin{equation}
u_{m_{h}}=\lambda ^{2}t_{m_{h}},\,\tau _{m_{h}}=t_{m_{h}}-t_{m_{h}^{\prime }}
\tag{3.18}
\end{equation}
This gives 
\begin{equation*}
<\Psi _{R},\prod_{j=1}^{2n}c_{T_{j},\lambda }^{\epsilon _{j}}(g_{j})\Psi
_{R}>=
\end{equation*}
\begin{equation*}
\sum_{\{M^{\prime }\equiv M^{c}:m_{n}^{\prime }<m_{h}\forall
\,h\}}\prod_{h=1}^{n}\biggl\{\int_{0}^{T_{m_{h}}}du_{m_{h}}%
\int_{-u_{m_{h}}}^{(T_{{m^{\prime }}_{h}}-u_{m_{h}})/\lambda
^{2}}dv_{m_{h}}\int dk_{m_{h}}
\end{equation*}
\begin{equation*}
\overline{g}_{{m^{\prime }}_{h}}(k_{m_{h}})g_{m_{h}}(k_{m_{h}})\exp
\{i\omega (k_{m_{h}})v_{m_{h}}\}\biggr\}\,W(-{\frac{1}{m}}%
\sum_{h=1}^{n}v_{m_{h}}k_{m_{h}},0)
\end{equation*}
\begin{equation*}
\times \exp \biggr\{{\frac{-i\hbar }{2m}}\sum_{h=1}^{n}\{\sum_{\alpha }^{{%
m^{\prime }}_{\alpha }<m_{h}}k_{m_{\alpha }}.k_{m_{h}}v_{m_{h}}-\sum_{\beta
}^{m_{\beta }<m_{h}}k_{m_{\beta }}.k_{m_{h}}v_{m_{h}}
\end{equation*}
\begin{equation*}
+\sum_{\alpha }^{m_{h}<{m^{\prime }}_{\alpha }<{m^{\prime }}%
_{h}}k_{m_{\alpha }}.k_{m_{h}}(v_{m_{\alpha }}-v_{m_{h}}+(u_{m_{\alpha
}}-u_{m_{h}})/\lambda ^{2})
\end{equation*}
\begin{equation}
-\sum_{\beta }^{m_{h}<m_{\beta }<{m^{\prime }}_{h}}k_{m_{\beta
}}.k_{m_{h}}(-v_{m_{h}}+(u_{m_{\beta }}-u_{m_{h}})/\lambda
^{2})-|k_{m_{h}}|^{2}v_{m_{h}}\}\biggr\}.  \tag{$3.19$}
\end{equation}
By an application of the Riemann-Lebesgue lemma, we have that the
oscillatory factors of the type $e^{ik^{2}u/\lambda ^{2}}$ cause the
associated term to vanish in the limit $\lambda \rightarrow 0$. By examining
the phase in (3.19) we see that, for each fixed $h=1,...,n$ and for any $%
\alpha $ 
\begin{equation}
m_{h}<m_{\alpha }<m_{h}^{\prime }\Leftrightarrow m_{h}<m_{\alpha }^{\prime
}<m_{h}^{\prime }  \tag{$3.20$}
\end{equation}
but this only possible for the Wigner partition. Hence only $M^{\prime }=%
\overline{M}$ survives the limit. Only in this case is the phase term
independent of $u_{m_{h}},h=1,...,n$ and explicitly it equals 
\begin{equation*}
\exp \biggr\{{\frac{-i\hbar }{2m}}\sum_{h=1}^{n}\{\sum_{\alpha }^{\overline{m%
}_{\alpha }<m_{h}}k_{m_{\alpha }}.k_{m_{h}}v_{m_{h}}-\sum_{\alpha
}^{m_{\alpha }<m_{h}}k_{m_{\alpha }}.k_{m_{h}}v_{m_{h}}
\end{equation*}
\begin{equation}
-\sum_{\alpha }^{m_{h}<m_{\alpha }<\overline{m}_{\alpha }<\overline{m}%
_{h}}k_{m_{\alpha }}.k_{m_{h}}v_{m_{\alpha }}-|k_{m_{h}}|^{2}\}\biggr\}. 
\tag{$3.21$}
\end{equation}
the first three terms can be combined to read as 
\begin{equation*}
{\frac{-i\hbar }{2m}}\sum_{h,\alpha =1}^{n}\biggl\{\chi _{(\overline{m}%
_{\alpha },n]}(m_{h})-\chi _{(m_{\alpha },n]}(m_{h})-\chi _{(m_{\alpha },%
\overline{m}_{\alpha })}(m_{h})\biggr\}k_{m_{\alpha }}.k_{m_{h}}v_{m_{\alpha
}}
\end{equation*}
\begin{equation}
={\frac{i\hbar }{m}}\sum_{h,\alpha =1}^{n}k_{m_{\alpha
}}.k_{m_{h}}v_{m_{\alpha }}\chi _{(m_{\alpha },\overline{m}_{\alpha
})}(m_{h}),  \tag{$3.22$}
\end{equation}
where we reversed the roles of $\alpha $ and $h$ in the third term. The
final term is then just 
\begin{equation}
\exp \{{\frac{i\hbar }{2m}}\sum_{h=1}^{n}|k_{m_{h}}|^{2}v_{m_{h}}\}. 
\tag{$3.23$}
\end{equation}
It is now evident that the $2n$-point function takes the form indicated in
the statement of the theorem.
\end{proof}

We remark that the form of the correlation functions can be simplified. Let $%
l_{m_{h}}$ denote the particle's momentum after the emission vertex $%
C^{\dagger }(g_{m_{h}},t_{m_{h}})$, by momentum conservation we have 
\begin{equation}
l_{m_{h}}=p-\hbar \sum_{r}^{m_{r}<m_{h}<\overline{m}_{h}<\overline{m}%
_{r}}k_{m_{r}},  \tag{$3.24$}
\end{equation}
that is $l_{m_{h}}$ equals the incoming free momentum $p$ minus the sum of
all emitted but not yet reabsorbed reservoir quanta momenta: as the
structure is non-crossing this means that we sum over reservoir quanta with
momentum $k_{m_{r}}$ which have been emitted before the vertex $m_{r}<m_{h}$
but not yet reabsorbed $\overline{m}_{r}>m_{h}$ (and so $\overline{m}_{r}>%
\overline{m}_{h}$). Let $\hbar \Delta (l,k)$ be the energy violation
associated with each vertex, that is 
\begin{equation*}
\hbar \Delta (l,k):={\frac{1}{2m}}|l-\hbar k|^{2}+\hbar \omega (k)-{\frac{1}{%
2m}}|l|^{2}
\end{equation*}
\begin{equation}
\Rightarrow \Delta (l,k)=-{\frac{1}{m}}l.k+\omega (k)+{\frac{\hbar }{2m}}%
|k|^{2}.  \tag{$3.25$}
\end{equation}
Then we have 
\begin{equation*}
\langle \prod_{j=1}^{2n}C^{\epsilon _{j}}(g_{j},T_{j})\rangle
=\prod_{h=1}^{n}(T_{\overline{m}_{j}}\wedge T_{m_{j}})\,\int_{-\infty
}^{+\infty }d\tau _{m_{n}}...\int_{-\infty }^{+\infty }d\tau _{m_{1}}\int
dk_{m_{n}}...\int dk_{m_{1}}
\end{equation*}
\begin{equation}
\times \prod_{r=1}^{n}\left\{ \overline{g}_{\overline{m}%
_{r}}(k_{m_{r}})g_{m_{r}}(k_{m_{r}})\exp \{i\Delta (l_{m_{r}},k_{m_{r}})\tau
_{m_{r}}\}\right\} .  \tag{$3.26$}
\end{equation}

\section{ Interacting Fock Space}

The theory of Interacting Fock Space was developed in [6,9]. We give a
slightly different presentation of it in this section. Let $\mathcal{K}%
\subset \mathcal{L}^{2}(\mathbb{R}^{3})$ denote the subspace of Schwartz
functions such that for all $f,g\in \mathcal{K}$ one has 
\begin{equation}
\int_{-\infty }^{+\infty }dt\,|<f,e^{i\Omega t}g>|<\infty ,  \tag{$4.1$}
\end{equation}
where $\Omega $ denotes multiplication by $\omega (k)$ on $\mathcal{L}^{2}(%
\mathbb{R}^{3})$. For $f,g\in \mathcal{K}$ we have shown 
\begin{equation}
<C(f,t)C^{\dagger }(g,s)>=t\wedge s\,(f|g)  \tag{4.2}
\end{equation}
where 
\begin{equation}
(f|g)\equiv (f|g)_{p}=\int_{-\infty }^{+\infty }d\tau \int dk\,\overline{f}%
(k)g(k)\,e^{i\Delta (p,k)\tau }.  \tag{$4.3$}
\end{equation}
Now $(f|g)$ is an element of $\mathcal{P}$, the (commutative) $C^{\ast }$%
-algebra generated by $\{e^{ix.p}:p\in \mathbb{R}^{3}\}$. The subscript $p$
shall not be displayed in general. We shall denote by $\mathcal{K}_{\mathcal{%
P}}$ the $\mathcal{P}$-right-linear span of $\mathcal{K}$ and $\mathcal{L}_{%
\mathcal{P}}^{2}(\mathbb{R}^{3},\mathcal{K})$ the algebraic tensor product
of $\mathcal{L}^{2}(\mathbb{R}^{3})$ and $\mathcal{K}_{\mathcal{P}}$. The
two point function suggests that we study the bilinear form $(.|.):\mathcal{L%
}_{\mathcal{P}}^{2}(\mathbb{R}^{3},\mathcal{K})\times \mathcal{L}_{\mathcal{P%
}}^{2}(\mathbb{R}^{3},\mathcal{K})\mapsto \mathcal{P}$ defined by 
\begin{equation}
(\alpha \underline{\otimes }f|\beta \underline{\otimes }g):=<\alpha ,\beta
>_{\mathcal{L}^{2}(\mathbb{R})}(f|g).  \tag{$4.4$}
\end{equation}
Next we wish to construct an $n$-particle space over $\mathcal{L}_{\mathcal{P%
}}^{2}(\mathbb{R}^{3},\mathcal{K})$ using the $2n$-point function to define
the $n$-fold inner product. That is, construct $\odot ^{n}\mathcal{L}_{%
\mathcal{P}}^{2}(\mathbb{R}^{3},\mathcal{K})$ out of $\otimes ^{n}\mathcal{L}%
_{\mathcal{P}}^{2}(\mathbb{R}^{3},\mathcal{K})$ with 
\begin{equation*}
(\,(\chi _{\lbrack 0,t_{1}]}\underline{\otimes }f_{1})\odot ...\odot (\chi
_{\lbrack 0,t_{n}]}\underline{\otimes }f_{n})\,|\,(\chi _{\lbrack 0,s_{1}]}%
\underline{\otimes }g_{1})\odot ...\odot (\chi _{\lbrack 0,s_{n}]}\underline{%
\otimes }g_{n}\,):=
\end{equation*}
\begin{equation}
<C(f_{n},t_{n})...C(f_{1},t_{1})C^{\dagger }(g_{1},s_{1})...C^{\dagger
}(g_{n},s_{n})>.  \tag{$4.5$}
\end{equation}
The above $2n$-point function corresponds to the complete \textit{rainbow}
diagram: it equals 
\begin{equation*}
(s_{1}\wedge t_{1})...(s_{n}\wedge t_{n})\int_{-\infty }^{\infty }d\tau
_{1}...\int_{-\infty }^{\infty }d\tau _{n}\int dk_{1}...\int dk_{n}\,%
\overline{f_{1}}(k_{1})g_{1}(k_{1})...\overline{f_{n}}(k_{n})g_{n}(k_{n})
\end{equation*}
\begin{equation}
e^{i\Delta (p,k_{n})\tau _{n}}e^{i\Delta (p-k_{n},k_{n-1})\tau
_{n-1}}...e^{i\Delta (p-k_{n}-k_{n-1}-...-k_{2},k_{1})\tau _{1}}. 
\tag{$4.6$}
\end{equation}
However, introducing the transform 
\begin{equation}
F=F(k)\mapsto {\tilde{F}}={\tilde{F}}_{p}:=\int_{-\infty }^{+\infty }d\tau
\int dk\,F(k)\,e^{i\Delta (p,k)\tau }  \tag{$4.7$}
\end{equation}
and the convolution 
\begin{equation*}
{\tilde{G}}\ast {\tilde{F}}={\tilde{G}}\ast {\tilde{F}}_{p}:=\int_{-\infty
}^{+\infty }d\tau \int dk\,{\tilde{G}}_{p-k}F(k)\,e^{i\Delta (p,k)\tau }
\end{equation*}
\begin{equation}
=\int_{-\infty }^{\infty }d\tau _{1}\int_{-\infty }^{\infty }d\tau _{2}\int
dk_{1}\int dk_{2}\,G(k_{1})F(k_{2})e^{i\Delta (p,k_{2})\tau _{2}}e^{1\Delta
(p-k_{2},k_{1})\tau _{1}}  \tag{$4.8$}
\end{equation}
we can write the correlator more succinctly as 
\begin{equation}
(s_{1}\wedge t_{1})...(s_{n}\wedge t_{n})\,(f_{1}|g_{1})\ast ...\ast
(f_{n}|g_{n})_{p}.  \tag{$4.9$}
\end{equation}
Note that the repeated application of the convolution is not associative and
we shall always understand the (inductively defined) convention 
\begin{equation}
{\tilde{F}_{1}}\ast {\tilde{F}_{2}}\ast ...\ast {\tilde{F}_{n}}:=[{\tilde{F}%
_{1}}\ast {\tilde{F}_{2}}\ast ...]\ast {\tilde{F}_{n}}.  \tag{$4.10$}
\end{equation}
We are therefore lead to the identification 
\begin{equation*}
(\,(\alpha _{1}\underline{\otimes }f_{1})\odot ...\odot (\alpha _{n}%
\underline{\otimes }f_{n})\,|\,(\beta _{1}\underline{\otimes }g_{1})\odot
...\odot (\beta _{n}\underline{\otimes }g_{n})\,):=
\end{equation*}
\begin{equation}
<\alpha _{1},\beta _{1}>_{\mathcal{L}^{2}(\mathbb{R})}...<\alpha _{n},\beta
_{n}>_{\mathcal{L}^{2}(\mathbb{R})}\,(f_{1}|g_{1})\ast ...\ast (f_{n}|g_{n}),
\tag{$4.11$}
\end{equation}
or on absorbing the $\mathcal{L}^{2}(\mathbb{R})$ term 
\begin{equation}
(\phi _{1}\odot ...\odot \phi _{n}|\psi _{1}\odot ...\odot \psi _{n}):=(\phi
_{1}|\psi _{1})\ast ...\ast (\phi _{n}|\psi _{n}).  \tag{$4.12$}
\end{equation}
The inner product $(.|.)$ on $\odot ^{n}\mathcal{L}_{\mathcal{P}}^{2}(%
\mathbb{R}^{3},\mathcal{K})$ does not factor, as in the case with $\otimes
^{n}\mathcal{L}_{\mathcal{P}}^{2}(\mathbb{R}^{3},\mathcal{K})$, and for this
reason we refer to $\odot ^{n}\mathcal{L}_{\mathcal{P}}^{2}(\mathbb{R}^{3},%
\mathcal{K})$ as the \emph{interacting }$n$\emph{-particle space.}

So now we have two notions of product on $\mathcal{P}$: the ordinary $%
C^{\ast }$-algebra product and now the non-associative convolution $\ast $.
Likewise, in addition to the usual module product, we can introduce a new
product $\underline{\ast }:\mathcal{P}\times \mathcal{L}_{\mathcal{P}}^{2}(%
\mathbb{R}^{3},\mathcal{K})\mapsto \mathcal{L}_{\mathcal{P}}^{2}(\mathbb{R}%
^{3},\mathcal{K})$ having the property that, for all $c,b\in \mathcal{P}$
and $f,g\in \mathcal{L}_{\mathcal{P}}^{2}(\mathbb{R}^{3},\mathcal{K})$, 
\begin{equation}
(f|c\underline{\ast }g)=c\ast (f|g)  \tag{$4.13$}
\end{equation}
and 
\begin{equation}
b\underline{\ast }c\underline{\ast }g=(bc)\underline{\ast }g.  \tag{$4.14$}
\end{equation}
The mapping $:c\mapsto (c\underline{\ast }.)$ defines a module homomorphism
from $\mathcal{P}$ to $\mathcal{B}(\mathcal{L}_{\mathcal{P}}^{2}(\mathbb{R}%
^{3},\mathcal{K}))$. We note that we can write 
\begin{equation}
(\phi _{1}\odot \phi _{2}|\psi _{1}\odot \psi _{2})=(\phi _{1}|\psi )\ast
(\phi _{2}|\psi _{2})=(\phi _{2}|\,(\phi _{1}|\psi _{1})\underline{\ast }%
\psi _{2})  \tag{$4.15$}
\end{equation}
and by induction 
\begin{equation*}
(\phi _{1}\odot ...\odot \phi _{n}|\psi _{1}\odot ...\odot \psi _{n})=
\end{equation*}
\begin{equation}
(\phi _{n}|(\phi _{n-1}|...(\phi _{2}|(\phi _{1}|\psi _{1})\underline{\ast }%
\psi _{2})...\underline{\ast }\psi _{n-1})\underline{\ast }\psi _{n}). 
\tag{$4.16$}
\end{equation}

The \emph{interacting Fock space} is then defined as 
\begin{equation}
\Gamma _{I}(\mathcal{L}_{\mathcal{P}}^{2}(\mathbb{R}^{3},\mathcal{K}%
)):=\bigoplus_{n=0}^{\infty }(\odot ^{n}\mathcal{L}_{\mathcal{P}}^{2}(%
\mathbb{R}^{3},\mathcal{K})),  \tag{$4.17$}
\end{equation}
where we take $\odot ^{0}\mathcal{L}_{\mathcal{P}}^{2}(\mathbb{R}^{3},%
\mathcal{K})=\mathcal{P}$.

The creation operator $A^{\dagger }(\phi )$, $\phi \in \mathcal{L}_{\mathcal{%
P}}^{2}(\mathbb{R}^{3},\mathcal{K})$, on $\Gamma _{I}(\mathcal{L}_{\mathcal{P%
}}^{2}(\mathbb{R}^{3},\mathcal{K}))$ is then defined by 
\begin{equation*}
A^{\dagger }(\phi ):\odot ^{n}\mathcal{L}_{\mathcal{P}}^{2}(\mathbb{R}^{3},%
\mathcal{K})\mapsto \odot ^{n+1}\mathcal{L}_{\mathcal{P}}^{2}(\mathbb{R}^{3},%
\mathcal{K})
\end{equation*}
\begin{equation}
:\psi _{1}\odot ...\odot \psi _{n}\mapsto \phi \odot \psi _{1}\odot ...\odot
\psi _{n}.  \tag{$4.18$}
\end{equation}
Its formal adjoint is denoted $A(\phi )$ and we see 
\begin{equation*}
(\phi _{1}\odot ...\odot \phi _{n-1}|\,A(\phi )\,\psi _{1}\odot ...\odot
\psi _{n})=(\phi \odot \phi _{1}\odot ...\odot \phi _{n-1}|\psi _{1}\odot
...\psi _{n})
\end{equation*}
\begin{equation}
=(\phi _{n-1}|...(\phi _{1}|(\phi |\psi _{1})\underline{\ast }\psi _{2})...%
\underline{\ast }\psi _{n}).  \tag{$4.19$}
\end{equation}
As a result we may write the action of the annihilator as 
\begin{equation*}
A^{\dagger }(\phi ):\odot ^{n}\mathcal{L}_{\mathcal{P}}^{2}(\mathbb{R}^{3},%
\mathcal{K})\mapsto \odot ^{n-1}\mathcal{L}_{\mathcal{P}}^{2}(\mathbb{R}^{3},%
\mathcal{K})
\end{equation*}
\begin{equation}
:\psi _{1}\odot ...\odot \psi _{n}\mapsto (\phi |\psi _{1})\underline{\ast }%
\psi _{2}\odot ...\odot \psi _{n}.  \tag{$4.20$}
\end{equation}

To complete our construction of the noise space we need to specify the
state; this will just be the expectation in the vacuum state $\Phi $ given
by 
\begin{equation}
\Phi =1_{\mathcal{P}}\oplus 0\oplus 0\oplus 0...  \tag{$4.21$}
\end{equation}
For example, we have the four-point functions 
\begin{equation}
(\Phi |A(\phi _{4})A(\phi _{3})A^{\dagger }(\phi _{2})A^{\dagger }(\phi
_{1})\Phi )=(\phi _{3}\odot \phi _{4}|\phi _{2}\odot \phi _{1})=(\phi
_{3}|\phi _{2})\ast (\phi _{4}|\phi _{1})  \tag{$4.22$}
\end{equation}
and 
\begin{equation}
(\Phi |A(\phi _{4})A^{\dagger }(\phi _{3})A(\phi _{2})A^{\dagger }(\phi
_{1})\Phi )=(\phi _{4}|\phi _{3})(\phi _{2}|\phi _{1}).  \tag{$4.23$}
\end{equation}
The second one is easily computed once one realizes that $(\Theta |A(\phi
)A^{\dagger }(\psi )\Phi )=(\phi |\psi )(\Theta |\Phi )$ for all $\Theta \in
\Gamma _{I}(\mathcal{L}_{\mathcal{P}}^{2}(\mathbb{R}^{3},\mathcal{K}))$.

The limit operators $C^{\sharp }(g,t)$ are then described mathematically by 
\begin{equation}
C^{\sharp }(g,t):=A^{\sharp }(\chi _{\lbrack 0,t]}\underline{\otimes }g), 
\tag{$4.24$}
\end{equation}
with expectation given by $<.>=(\Phi |\,.\,\Phi )$. One easily sees that the
correlators, to all orders, are given by this prescription.

\section{The Interacting-Free Stochastic Sector of Quantum Field Theory}

Define an operator $A^{\dagger }(\alpha \underline{\otimes }f)$, for $\alpha
\in \mathcal{L}^{2}(\mathbb{R})$ and $f\in \mathcal{K}$, by 
\begin{equation}
A^{\dagger }(\alpha \underline{\otimes }f):=|\alpha >\underline{\otimes }%
\int_{-\infty }^{\infty }d\tau \int dk\,f(k)e^{i\omega (k)\tau }W(-{\frac{%
\tau }{m}}k,-k)\otimes \alpha ^{\dagger }(\tau ,k)  \tag{$5.1$}
\end{equation}
with adjoint 
\begin{equation}
A(\alpha \underline{\otimes }f)=<\alpha |\underline{\otimes }\int_{-\infty
}^{\infty }d\tau \int dk\,\overline{f}(k)e^{-i\omega (k)\tau }W({\frac{\tau 
}{m}}k,k)\otimes \alpha (\tau ,k)  \tag{$5.2$}
\end{equation}
where the operators $\alpha ^{\sharp }$ satisfy the scaled free relations 
\begin{equation}
\alpha (\tau ,k)\alpha (\tau ^{\prime },k^{\prime })=\delta (\tau ^{\prime
}-2\tau )\delta (k-k^{\prime }).  \tag{$5.3$}
\end{equation}
Let $\Phi $ denote the vacuum state 
\begin{equation}
\alpha (\tau ,k)\Phi =0.  \tag{$5.4$}
\end{equation}
The two-point functions are given by 
\begin{equation*}
<\Phi ,A(\alpha _{2}\underline{\otimes }f_{2})A^{\dagger }(\alpha _{1}%
\underline{\otimes }f_{1})\Phi >=
\end{equation*}
\begin{equation*}
<\alpha _{2}|\,.\,|\alpha _{1}>\int_{-\infty }^{\infty }d\tau _{2}\int
dk_{2}\int_{-\infty }^{\infty }d\tau _{1}\int dk_{1}\,\overline{f}%
_{2}(k_{2})e^{-i\omega (k_{2})\tau _{2}}f_{1}(k_{1})e^{i\omega (k_{1})\tau
_{1}}
\end{equation*}
\begin{equation*}
\times W({\frac{\tau _{2}}{m}}k_{2},k_{2})W(-{\frac{\tau _{1}}{m}}%
k_{1},-k_{1})\delta (k_{1}-k_{2})\delta (\tau _{1}-2\tau _{2})
\end{equation*}
\begin{equation*}
=<\alpha _{2},\alpha _{1}>\int_{-\infty }^{\infty }d\tau \int dk\,\overline{f%
}_{2}(k)e^{-i\omega (k)\tau }f_{1}(k)e^{i\omega (k)2\tau }\,W({\frac{\tau }{m%
}}k,k)W(-{\frac{2\tau }{m}}k,-k)
\end{equation*}
\begin{equation*}
=<\alpha _{2},\alpha _{1}>\int_{-\infty }^{\infty }d\tau \int dk\,\overline{f%
}_{2}(k)f_{1}(k)e^{i\omega (k)\tau }W(-{\frac{\tau }{m}}k,0)e^{{\frac{i\hbar
\tau }{2m}}|k|^{2}}
\end{equation*}
\begin{equation}
\equiv <\alpha _{2},\alpha _{1}>(f_{2}|f_{1})  \tag{$5.5$}
\end{equation}
The four point functions are also easily obtained 
\begin{equation*}
<\Phi ,A(\alpha _{4}\underline{\otimes }f_{4})A(\alpha _{3}\underline{%
\otimes }f_{3})A^{\dagger }(\alpha _{2}\underline{\otimes }f_{2})A^{\dagger
}(\alpha _{1}\underline{\otimes }f_{1})\Phi >=
\end{equation*}
\begin{equation*}
<\alpha _{4}|\,.<\alpha _{3}|\,.\,|\alpha _{2}>\,|\alpha _{1}>\int_{-\infty
}^{\infty }d\tau _{4}\int dk_{4}\int_{-\infty }^{\infty }d\tau _{1}\int
dk_{1}
\end{equation*}
\begin{equation*}
\times \overline{f}_{4}(k_{4})e^{-i\omega (k_{4})\tau
_{4}}f_{1}(k_{1})e^{i\omega (k_{1})\tau _{1}}\int_{-\infty }^{+\infty }d\tau
\int dk\,\overline{f}_{3}(k)f_{2}(k)e^{i\omega (k)\tau }e^{{\frac{i\hbar
\tau }{2m}}|k|^{2}}
\end{equation*}
\begin{equation*}
\times W({\frac{\tau _{4}}{m}}k_{4},k_{4})W(-{\frac{\tau }{m}}k,-k)W(-{\frac{%
\tau _{1}}{m}}k_{1},-k_{1})\delta (k_{1}-k_{4})\delta (\tau _{1}-2\tau _{4})
\end{equation*}
\begin{equation*}
=<\alpha _{4},\alpha _{1}><\alpha _{3},\alpha _{2}>\int_{-\infty }^{\infty
}d\tau ^{\prime }\int dk^{\prime }\int_{-\infty }^{\infty }d\tau \int dk\,%
\overline{f}_{4}(k^{\prime })f_{1}(k^{\prime })\overline{f}%
_{3}(k)f_{2}(k)e^{i\omega (k)2\tau }
\end{equation*}
\begin{equation*}
\times W(-{\frac{\tau }{m}}k-{\frac{\tau ^{\prime }}{m}}k^{\prime },0)e^{{%
\frac{i\hbar }{2m}}|k^{\prime }|^{2}\tau ^{\prime }}e^{{\frac{i\hbar }{m}}%
k.k^{\prime }\tau }
\end{equation*}
\begin{equation}
=<\alpha _{4},\alpha _{1}><\alpha _{3},\alpha _{2}>(f_{3}|f_{2})\ast
(f_{4}|f_{1}).  \tag{$5.6$}
\end{equation}
It is clear that the operators $A^{\dagger }(\alpha \underline{\otimes }f)$
defined above reproduce the same correlations as those introduced in the
last section.

\bigskip

\noindent $\underline{\mathrm{Acknowledgements:}}$ The author is very happy
to express gratitude to Prof. Luigi Accardi for many inspiring conversations
and for the warm hospitality afforded to him while writing this paper at the
Centro Vito Volterra. Thanks also goes to Prof. Yun Gang Lu for giving many
insights into theory of Hilbert Modules.

\bigskip

[1] R.P. Feynman, A.R. Hibbs. \textit{Quantum Mechanics and Path Integrals.}
Mc Graw - Hill Inc., (1965)

\bigskip

[2] L.Accardi, A.Frigerio and Y.G.Lu. \textit{The Weak Coupling Limit as a
Quantum Functional Central Limit Theorem.} Comm.Math.Phys.131, 537-570
(1990).

\bigskip

[3] L.Accardi, Y.G.Lu and I.Volovich. \textit{The Stochastic Sector of
Quantum Field Theory} Mathematicheskie Zametki (1994)

\bigskip

[4] L.Accardi, A.Frigerio and Y.G.Lu. \textit{The Weak Coupling Limit
Without Rotating Wave Approximation.} Ann.Inst,H.Poincar\'{e} Phys.Th. Vol.
54, No 4 (1991)

\bigskip

[5] L.Accardi, J.Gough and Y.G.Lu. \textit{On the Stochastic Limit For
Quantum Theory.} Rep.on Math.Phys. Vol.36, No. 2/3 (1995).

\bigskip

[6] L.Accardi and Y.G.Lu. \textit{The Wigner Semi-circle Law in Quantum } 
\textit{Electro-} \textit{Dynamics.} Preprint Centro Volterra 132. (January
93). L.Accardi and Y.G.Lu. \textit{Quantum Electro-Dynamics: The Master And
Langevin Equation}, Mathematical Approach to Fluctuations. T.Hida (ed.)
World Scientific (1994)

\bigskip

[7] D.Voiculescu. \textit{Free Non-Commutative Random Variables, Random
Matrices and the }$II_{1}$\textit{\ Factors of Free Groups}. Quantum
Probabitity and Related Fields. 473-487, (1991)

\bigskip

[8] B.K\"{u}mmerer, R.Speicher. \textit{Stochastic Integration on the Cuntz
Algebra }$O_{\infty }$. Jour. Funct. Anal. 103, No2 ,372-408 (1992),
R.Speicher. {A new example of Independence and White Noise}. Prob. Th. Rel.
Fields. 84, 141-159, (1990)

\bigskip

[9] Y.G.Lu , \textit{Quantum Stochastic Calculus on the Interacting Free
Fock Space }. Submitted to Journ. Funct. Anal.

\end{document}